# Was ist eine Professur für Künstliche Intelligenz?

**Prof. Dr. Kristian Kersting**[a,c], **Prof. Jan Peters, PhD.**[a,c], **Prof. Rothkopf, PhD.**[b,c]

kersting@cs.tu-darmstadt.de, peters@ias.informatik.tu-darmstadt.de, rothkopf@psychologie.tu-darmstadt.de

[a]Fachbereich Informatik, [b]Fachbereich Humanwissenschaften,
[c]Center for Cognitive Science, TU Darmstadt, Darmstadt, Deutschland

17. Februar 2019

***English Abstract:*** *The Federal Government of Germany aims to boost the research in the field of Artificial Intelligence (AI). For instance, 100 new professorships are said to be established. However, the white paper of the government does not answer what an AI professorship is at all. In order to give colleagues, politicians, and citizens an idea, we present a view that is often followed when appointing professors for AI at German and international universities. We hope that it will help to establish a guideline with internationally accepted measures and thus make the public debate more informed.*

***Zusammenfassung:*** *Die deutsche Bundesregierung will die Forschung in der Künstlichen Intelligenz in Deutschland deutlich mehr fördern als bisher. Es sollen z.B. 100 neue Professuren entstehen. Allerdings beantwortet das Strategiepapier nicht, was eine Professur für Künstliche Intelligenz überhaupt ist. Um Kollegen, Politikern und Bürgern eine Idee zu geben, stellen wir eine Einordnung vor, wie sie in Berufungsverfahren an deutschen und internationalen Universitäten üblich ist. Wir hoffen, dass eine solche Einordnung einen Beitrag zu einem Leitfaden mit Messpunkten liefert und so die öffentliche Debatte aufgeklärter gestaltet.*

### Das wissenschaftliche Gebiet der Künstliche Intelligenz (KI)

Zweifellos erlebt die deutsche KI Landschaft derzeit einen grundlegenden Wandel. An der aktuellen Diskussion um KI und die KI Strategie der Bundesregierung[1] beteiligen sich viele KI-Experten. Aber was macht eine Person zu einer KI-Expertin?

Im Allgemeinen sind Expertinnen Personen, "die sich — ausgehend von spezifischem Praxis- oder Erfahrungswissen, das sich auf einen klar begrenzbaren Problemkreis bezieht — die Möglichkeit geschaffen haben, mit ihren Deutungen das konkrete Handlungsfeld sinnhaft und handlungsleitend zu strukturieren"[2]. An den Universitäten sind Experten typischerweise Professorinnen und Professoren, die im Sinne des humboldtschen Bildungsideals eigenverantwortliche wissenschaftlicher Forschung und Lehre durchführen. Um nun zu verstehen, was eine KI Professur ist, müssen wir erst einmal klären, was KI ist.

Für viele Forscher im Feld ist der Dartmouth Workshop von 1956 die Geburtsstunde der KI. Hier versammelten sich über einen Zeitraum von acht Wochen in verschiedenen Konstellationen unter anderem John McCarthy, Claude Shannon, Warren McCulloch, Marvin Minsky, John Nash, Herbert Simon und Allen Newell. Diese Liste zeigt dabei die von Anfang an erkannte Notwendigkeit, Experten in der Quantifizierung und Formalisierung menschlicher Denkvorgänge, was heute als Kognitionswissenschaft gilt, einzubinden. Als thematische Liste wurde schon bei diesem Workshop Computer, Verarbeitung natürlicher Sprache, Neuronale Netzwerke, Theorie der Computation, Abstraktion und Kreativität genannt, welche noch heute relevant sind. Somit bezog sich der Begriff Künstliche Intelligenz von Anfang an unmittelbar auch auf menschliche Intelligenz, wobei es sich herausgestellt hat, dass eine klare Definition, was menschliche oder natürlich Intelligenz überhaupt sei, recht schwierig ist, und man mit dieser Unsicherheit leben muss, so wie in anderen akzeptierten Wissenschaftsdisziplinen wie z.B. der Psychologie, der Biologie oder der Ökonomie auch. Dementsprechend hat mit der Geburt der Künstliche Intelligenz auch eine weitere Wissenschaft das Licht der Welt erblickt, nämlich die Cognitive Science, die sich damit beschäftigt zu verstehen, was menschliche oder natürliche Intelligenz überhaupt ausmacht. So haben sich die beiden Felder der Künstlichen Intelligenz und der Cognitive Science immer wieder gegenseitig inspiriert und beeinflusst.

---

[1] https://www.bundesregierung.de/resource/blob/997532/1550276/3f7d3c41c6e05695741273e78b8039f2/2018-11-15-ki-strategie-data.pdf, Zugriff 16. Februar 2019

[2] Alexander Bogner, Beate Littig & Wolfgang Menz (Hrsg.) (2002). Das Experteninterview. Theorie, Methode, Anwendung, VS Verlag für Sozialwissenschaften, S. 45

Aber was ist denn nun KI? Im Zuge des Dartmouth Workshops führte John McCarthy den Begriff "Künstliche Intelligenz" (KI, engl. artificial intelligence) ein und definierte[3] ihn wie folgt:

> AI is "the science and engineering of making intelligent machines, especially intelligent computer programs. It is related to the similar task of using computers to understand human intelligence, but AI does not have to confine itself to methods that are biologically observable."

KI ist also die (Ingenieur)Wissenschaft intelligenter Maschinen, insbesondere intelligenter Computerprogramme. Das sind Algorithmen[4] — eindeutige Handlungsvorschriften zur Lösung eines Problems oder einer Klasse von Problemen und bei einer bestimmten Eingabe eine bestimmte Ausgabe überführen — des Problemlösens, Denkens und Lernens. Daher ist die KI-Forschung auch stark in der Informatik zu verankern.

Die Vision des Dartmouth Workshops prägt auch heute noch die moderne KI-Forschung. Das internationale Standardlehrbuch für Künstlichen Intelligenz (Russel und Norvig, Articial Intelligence: A modern Approach, 3. Auflage, Pearson 2009) behandelt:

**Problemlösung**
   *Suche, Constraint Programming, Satisfiability, ...*
   Ein wichtiger Aspekt der KI ist die zielorientierte Problemlösung. Die Lösung vieler Probleme (z.B. Tic-Tac-Toe, Stundenplan, Schach) kann beschrieben werden, indem man eine Reihe von Aktionen findet, die zu einem gewünschten Ziel führen. Jede Aktion ändert den Zustand und das Ziel ist es, die Reihenfolge der Aktionen und Zustände zu finden, die vom Anfangszustand bis zum Endzustand führen.

**Wissensrepräsentation & Schlussfolgern**
   *Logik, Schlussfolgern, Wissensrepräsentation, Planung, ...*
   Das Wissen über die Welt wird maschinenlesbar so dargestellt, dass eine Maschine es nutzen kann, um komplexe Aufgaben zu lösen wie z.B. die Diagnose einer Krankheit zu stellen oder einen Dialog in einer natürlichen Sprache zu führen

---

[3] http://jmc.stanford.edu/artificial-intelligence/what-is-ai/index.html Zugriff 16. Februar 2019
[4] https://de.wikipedia.org/wiki/Algorithmus, Zugriff 16. Februar 2019

| | |
|---|---|
| **Unsicherheit & Schlussfolgern** | |
| *Probabilistische Modelle, Entscheidungstheorie, …* | |
| Die Welt ist nicht sicher. Es gibt Situationen, bei denen der Eintritt von Ereignissen nicht mit Sicherheit festgestellt oder sogar vorausgesagt werden kann. Die Auswahl von möglichen Alternativen und deren Auswirkungen sind nicht vollständig bekannt. | |
| **Maschinelles Lernen** | |
| *(Un)überwachtes Lernen, Reinforcement Learning, Deep Learning, …* | |
| Eine Maschine lernt aus Beispielen und kann die erlernten Muster und das erlernte Wissen auf neue Beispiele verallgemeinern. Die Maschine lernt also aus Erfahrungen ihr zukünftiges Verhalten zu verbessern. | |
| **Wahrnehmung und Sehen** | |
| *Objekterkennung, -segmentierung, visuelle Frage/Antwort Aufgaben, …* | |
| Die Fähigkeit zur Verarbeitung visueller Information ist eine Grundbedingung für viele automatisierte Prozesse. Mithilfe von Regeln und Algorithmen werden Bilder und andere sensorischen Eingaben verarbeiten, interpretiert und auch generiert. | |
| **Verstehen und Generierung von natürlicher Sprache** | |
| *Parsen, Semantische Einbettungen, Frage/Antwort Aufgaben, …* | |
| Maschinen müssen natürliche Sprache erfassen, verarbeiten, verstehen und auch generieren können. Ziel ist es, eine möglichst weitreichende Kommunikation zwischen Mensch und Computer per Sprache zu schaffen. | |
| **Interaktion** | |
| *Multimodale und -mediale, benutzerorientierte Modellierung von Anwendungen* | |
| (Computer)programme müssen mit ihrer Umwelt und insbesondere mit uns Menschen interagieren. Ziel ist es, eine möglichst weitreichende Kommunikation zwischen Mensch und Computer zu schaffen. | |
| **Robotik** | |
| *Greifen von Objekten, Stehen, Laufen, Exploration der Umgebung, …* | |
| Design und Entwicklung von Robotern, die mittels KI-Algorithmen autonome mit der physischen Welt interagieren | |

Eine Professur für Künstliche Intelligenz und ihre Forschung sollte also diese und ähnliche Fragestellungen adressieren. Die Ergebnisse — mit der Zeit auch zu neuen Fragestellungen wie z.B. algorithmischer Fairness und Ethik — werden auf wissenschaftlichen Tagungen und in Zeitschriften zu dem Thema KI publiziert und diskutiert.

### Reputation in der KI-Forschung: Wissenschaftliche Publikationen

Die KI-Forschungslandschaft an Hochschulen zeichnet sich durch eine thematische und methodische Breite der Forschung und die Ausbildung des wissenschaftlichen Nachwuchses aus. Jede Wissenschaft „lebt von der Glaubwürdigkeit und ihrer internen Kontrolle", wie Andreas Hotho[5] betont. Und weiter, „Wissenschaftler haben in der Gesellschaft eine hohe Reputation, und wissenschaftliche Erkenntnisse sind insbesondere in Deutschland entsprechend anerkannt. Woher kommt dieses hohe Ansehen? Der wissenschaftliche Qualitätssicherungsprozess ist entsprechend ausgefeilt und versucht, mögliche Fehler frühzeitig und effektiv zu finden und zu verhindern. Erkenntnisse sind so zu präsentieren, dass Kollegen diese nachvollziehen und zugehörige Experimente wiederholen können. Im wissenschaftlichen Diskurs setzen sich die Wissenschaftler mit den Ergebnissen der Kollegen auseinander und entwickeln diese weiter. In der Praxis bedeutet dies im Idealfall, dass nach dem Schreiben eines wissenschaftlichen Aufsatzes dieser einen Begutachtungsprozess durchläuft. Nur Aufsätze, deren Ergebnisse laut Gutachter richtig und wichtig sind und die die wissenschaftlichen Erkenntnisse angemessen präsentieren, werden in den entsprechenden Zeitschriften veröffentlicht. Je renommierter die Zeitschrift, desto härter und zum Teil auch länger ist der Prüfungsprozess. Aber auch nach der Veröffentlichung erfahren die Ergebnisse stetig eine Weiterentwicklung und Kontrolle durch die wissenschaftliche Gemeinschaft. Sollten sich Ergebnisse als falsch herausstellen, dann werden sie idealerweise revidiert und korrigiert. Dies ist nicht ungewöhnlich, sondern der normale Fortschritt."

---

[5] Andreas Hotho. „Social Media und Künstliche Intelligenz in der Wissenschaftskommunikation: Ein visionärer Ausblick." In P. Weingart, H. Wormer, A. Wenninger, R.F. Hüttl (Hg.), „Perspektiven der Wissenschaftkommunikation im digitalen Zeitalter", Velbrück Wissenschaft, 2017, S. 280.

Dieses sogenannte Peer-Review[6] (englisch von Peer, Gleichrangiger und Review, Gutachten) wird auch in der KI-Forschung verfolgt. Allerdings sind in der KI-Forschung so wie in der Informatik allgemein Publikationen in Tagungsbänden mindestens so angesehen wie Veröffentlichungen in Zeitschriften. Die KI Wissenschaftlerinnen treffen sich auf internationalen Workshops und Konferenzen, tauschen sich aus und beurteilen ihre Arbeiten. Die Qualität der Begutachtungsprozesse, also die Reputation einer Tagung oder einer Zeitschrift kann z.B. durch das **CORE Ranking**[7] eingeschätzt werden. Natürlich gibt es andere Maße, die man heranziehen kann, wie z.B. **Google Scholars**[8] **h5-Index**. Allgemein ist aber anzumerken, dass keines der Maße perfekt ist und immer nur einen Eindruck liefern kann. Das CORE Ranking basiert auf freiwilligen Einschätzungen der internationalen Forschungs-gemeinschaft, die durch ein Komitee aus australischen Kollegen final eingeschätzt werden, und folgt im Wesentlichen dem US-amerikanischen Schulnotensystem, von A* für die zu erwartenden härtesten Begutachtungsprozesse, über A, B und C hin zu „nicht kategorisiert", weil die Tagung oder das Journal z.B. noch zu neu ist. Google Scholars h5-Index ist die größte Zahl h, so dass h Veröffentlichungen der letzten 5 Jahre jeweils mindestens h Zitierungen bekommen haben: je höher, desto besser.

Die zentralen KI-Tagungen und Zeitschriften, die jeweils alle Themen der KI abdecken, sind

| Zentrale KI-Tagungen und Zeitschriften, die alle Themen der KI behandeln |
|---|
| **CORE A*** |
| AAAI Conference on Artificial Intelligence (AAAI) **h5 69** |
| International Conference on Artificial Intelligence (IJCAI) **h5 61** |
| Artificial Intelligence Journal (AIJ) **h5 46** |
| **CORE A** |
| Journal of Artificial Intelligence Research (JAIR) **h5 37** |
| European Conference on Artificial Intelligence (ECAI) **h5 22** |

---

[6] https://de.wikipedia.org/wiki/Peer-Review

[7] http://www.core.edu.au/conference-portal Es erhebt keinen Anspruch auf Vollständigkeit. Zugriff 17. Februar 2019

[8] Google Scholar ermittelt Zitierungsstatistiken automatisch, bereinigt aber z.B. nicht Selbstzitierungen und erhebt keinen Anspruch auf Vollständigkeit. So kann man z.B. einen Eindruck für die Zitierungen einzelner Personen wie z.B. John McCarthy bekommen (https://scholar.google.com/citations?user=SuVID2wAAAAJ&hl=en) oder für Tagungen und Zeitschriften zu Themengebiete wie z.B. „neural, machine-learning, artificial-intelligence" (https://scholar.google.it/citations?hl=en&view_op=search_venues&vq=neural+OR+machine-learning+OR+artificial-intelligence+) Zugriff 17. Februar 2019

Natürlich gibt es auch andere KI-Tagungen und Zeitschriften, die alle Gebiete der KI diskutieren, wie z.B. International Symposium on Artificial Intelligence and Mathematics (ISAIM), IEEE Intelligent Systems oder Annals of Mathematics and Artificial Intelligence, aber diese liegen nicht so sehr im aktuellen Fokus der internationalen KI Forschungsgemeinschaft. Es gibt auch Publikationsorgane, die KI im Kontext von Anwendungsgebieten behandeln oder Grundlagen in der Informatik entwickeln, die für die KI wichtig sind, wie z.B. Artificial Intelligence in Medicine (AIME), Bioinformatics, ACM SIGMOD International Conference on Management of Data (SIGMOD). Diese beschäftigen sich aber nicht im Kern mit den Fragestellungen der KI im Sinne der Definition von John McCarthy. Daher werden wir sie nicht weiter zur Abgrenzung heranziehen.

### Der Forschungsanspruch einer Professur für Künstliche Intelligenz

Der Forschungsanspruch einer Professur für Künstliche Intelligenz sollte sein, dass ihre Ergebnisse in den internationalen KI-Tagungen und Zeitschriften — AAAI, AIJ, ECAI, IJCAI, JAIR — publiziert werden. Allerdings sind die Fragestellungen der KI nicht einfach und erfordern oftmals eine Spezialisierung. Daher haben sich mit der Zeit Teildisziplinen der KI mit eigenen Tagungen und Zeitschriften herausgebildet wie z.B. ACL, CVPR, ICAPS, ICML, JMLR, KR, NeurIPS, TPAMI oder UAI (eine genaue Auflistung erfolgt später). Eine KI-Professur kann und/oder sollte ihre Forschungsergebnisse auch hier publizieren. Das ist nicht ungewöhnlich. Viele KI-Professuren setzen einen Schwerpunkt in ihrer Forschung.

> Zusammenfassung: Die KI-Forschung behandelt Algorithmen zum Problemlösen, Denken, Lernen und Handeln im Allgemeinen. Die Ergebnisse werden in Tagungsbänden und Zeitschriften (AAAI, IJCAI, AIJ, JAIR, ECAI) veröffentlicht. Ihren Teildisziplinen behandeln nur eine dieser Aufgaben, z.B. das Maschinelle Lernen, Computersehen, etc. und haben oft eigene Tagungen und Zeitschriften.

Am einfachsten stellt man sich das als konzentrische Kreise vor: KI ist der äußere und ihre Teildisziplinen sind innere Kreis. Sie ist stark in der Informatik verankert. Ihre Zwillingsdisziplin ist die Cognitive Science und sie hatte enge Beziehungen zur Psychologie, den Neurowissenschaften und der Philosophie und anderen Disziplinen.

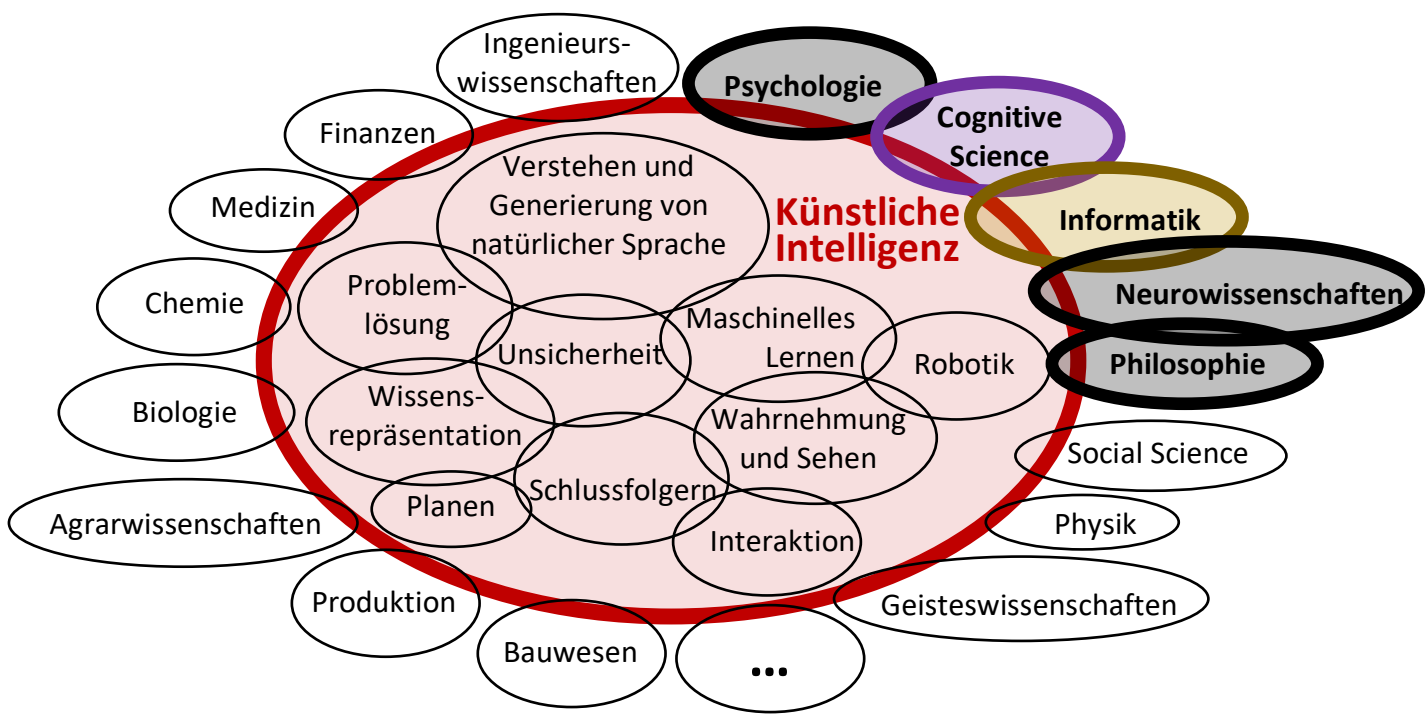

## Die Publikationslandschaft der KI

Nun zur Einordnung und Abgrenzung von KI-Professuren. Wir folgen hier dem Ideal einer Begutachtung innerhalb eines Berufungsverfahrens an einer Hochschule. Idealerweise sollten die Forschungsbeiträge einer KI-Professur auf den zentralen KI-Organen publiziert sein. Denn dann muss sich die Forschung und ihre Ergebnisse in den allgemeinen Kontext der KI einordnen und sich in Konkurrenz zu den anderen Teilgebieten behaupten. Sollte sich eine Professur nur mit einer Teildisziplin beschäftigen und auch nur in Spezial-Organen publizieren, so kann man diese zwar als eine Professur für KI einordnen, aber eher als eine Professur der Teildisziplin wie z.B. des Maschinelle Lernens. Man sollte dann von einer "Professur der Künstlichen Intelligenz für X" sprechen und den Bezug der Forschung zur Künstlichen Intelligenz prüfen.

Um den wissenschaftlichen Beitrag einer KI Forscherin zu ermessen, zieht man üblicherweise Online-Dienste wie **DBLP**[9] und **Google Scholar**[10] heran. Sie geben der Gutachterin gute Hinweise darauf, was, wieviel und wo jemand publiziert hat und wie viele Zitierung die Forschungsarbeit generiert hat. Wir möchten an dieser Stelle betonen, dass eine wissenschaftliche Leistung nicht allein aufgrund der Zahl der Publikationen bemessen werden sollte. Denn wie sagte schon Einstein zum Publikationsdruck: „Wenn sich jemand in seiner akademischen Laufbahn gezwungen sieht, eine große Anzahl wissenschaftlicher Schriften zu veröffentlichen, birgt das die Gefahr einer geistigen Verflachung."[11] Auch die Zahl der Zitierungen ist für sich allein genommen kein zuverlässiger Indikator sein. Die Einschätzung der wissenschaftlichen Leistung wird immer eine Mischung vieler Faktoren sein, die in einer Berufungskommission in den Händen anerkannter Wissenschaftlerinnen liegen muss — den Mitgliedern der Kommission und den externen Gutachterinnen.

Aber wie geht eine Gutachterin vor? Ziel eines Gutachtens ist, die Exzellenz der Forschung und die Expertise der Kandidatin einzuschätzen. Dabei werden viele Faktoren berücksichtigt wie z.B. erhaltene Forschungspreise, die Leitung von wissenschaftlichen Tagungen und (Verbund)projekten, Auslandserfahrungen, Erfahrungen im Hochschulmanagement, die Einwerbung von Forschungsgeldern, der Erfahrungen in der Lehre oder in der Wirtschaft, und die Tätigkeit als Gutachterin von internationalen KI-Tagungen und Zeitschriften. Der wesentliche Kern besteht aber aus den Forschungsleistungen, die man durch Publikationsleistungen versucht abzuschätzen: was, wieviel und wo hat die Kandidatin publiziert und wie viele Zitierung hat ihre Forschungsarbeit erzeugt? Dazu liest die Gutachterin die wichtigsten Publikationen einer Kandidatin, beurteil den wissenschaftlichen Beitrag der Veröffentlichungen und betrachtet das Renommee der Publikationsorte, deren Begutachtungsprozesse ja gerade der wissenschaftlichen Qualitätssicherung dienen.

Die CORE Rankings sowie die Google Scholars h5-Indizes der bereits genannten, zentralen KI Tagungen und Zeitschriften zeigen, dass sie international sehr angesehen sind.

---

[9] https://dblp.uni-trier.de/
[10] https://scholar.google.com/citations?user=SuVID2wAAAAJ&hl=en
[11] Isaacson, Walter: Einstein: His Life and Universe. New York: Simon & Schuster, 2007, S. 79. Zitiert in Geman, Donald und Stuart Geman: „Science in the Age of Selfies". PNAS 113:34 (23. August 2016), S. 9384-9387. Übernommen aus der Erklärung der drei Nationalen Akademien der Wissenschaften Académie des sciences, Leopoldina und Royal Society zu wissenschaftlichen Publikationen. https://www.leopoldina.org/fileadmin/redaktion/Publikationen/2016_Joint_Statement_on_scientific_publications_DE.pdf, Zugriff 15. Februar 2019

## Spezialisierte Tagungen und Zeitschriften der Künstlichen Intelligenz

Neben den bereits angeführten, zentralen Tagungen und Zeitschriften führen wir im Folgenden weitere wichtige[12] Tagungen und Zeitschriften auf, über die sich KI-Forscher austauschen. Die Tagungen und Zeitschriften selbst müssen sich aber nicht alleinig den Themen der KI widmen.

Wir erheben keinen Anspruch auf Vollständigkeit, die Liste sollte aber ein gutes Gefühl für relevante KI Tagungen und Zeitschriften geben. Die Tagungen und Zeitschriften sind wieder in Gruppen entsprechend ihres CORE Rankings und dann entsprechend des Google Scholar h5-Index sortiert. Dabei ist zu beachten, dass das Ranking der Journale aus dem Jahre 2010 stammt, so dass aktuelle Entwicklungen nicht abgedeckt sein müssen. Google Scholar erhebt auch keinen Anspruch auf Vollständigkeit. So können Tagungen und Zeitschriften einfach zu neu sein, um einen Google Scholar h5-Index zu haben.

## Problemlösen

Ein wichtiger Aspekt der KI ist die zielorientierte Problemlösung. Die Lösung vieler Probleme (z.B. Tic-Tac-Toe, Stundenplan, Schach) kann beschrieben werden, indem man eine Reihe von Aktionen findet, die zu einem gewünschten Ziel führen. Jede Aktion ändert den Zustand und das Ziel ist es, die Reihenfolge der Aktionen und Zustände zu finden, die vom Anfangszustand bis zum Endzustand führen.

| **Problemlösung** |
|---|
| Suche, Constraint Programming, Satisfiability, ... |
| **Core A*** |
| Principles of Knowledge Representation and Reasoning (KR) **h5 21** |
| **Core A** |
| Intern. Conference on Principles and Practice of Constraint Programming (CP) **h5 22** |
| Intern. Conference on Theory and Applications of Satisfiability Testing (SAT) **h5 21** |
| Constraints **h5 15** |
| **Core B** |
| International Conference on the Integration of Constraint Programming, AI, and Operations Research (CPAIOR) **h5 17** |
| Intern. Symposium Abstraction, Reformulation, and Approximation (SARA) **h5 -** |
| **Core Unranked** |
| International Symposium on Combinatorial Search (SOCS) **h5 17** |

---

[12] Wir haben mittels DBLP geprüft, welche andere Tagungen und Zeitschriften für Veröffentlichungen mit einem KI Bezug von Kollegen genutzt werden, die in den zentralen KI Tagungen und Zeitschriften präsent sind. Auch haben wir in CORE und Google Scholar nach KI Tagungen und Zeitschriften mittels KI Stichworten gesucht (z.B. „artificial-intelligence", „neural", „machine-learning", „combinatorial-search", …) und aufgrund unserer eigenen Diskussionen vereinzelt Venues wieder gestrichen, die keinen wissenschaftlichen Bezug hatten. Die Zuordnung von Tagungen und Zeitschriften zu einer KI Teildisziplinen ist nicht immer eindeutig, wir bitte das zu entschuldigen. Es wird auch kein Anspruch auf Vollständigkeit erhoben, aber die wichtigen nach CORE 2018 Ranking (http://portal.core.edu.au/conf-ranks/) sowie ERA2010 (http://portal.core.edu.au/jnl-ranks/) sind enthalten. Auch die KI Tagungen, die vom AI Index Report 2018 (https://aiindex.org/) als wichtig angesehen werden, sind enthalten. Der AI Index wurde im Rahmen der „One Hundred Year Study on AI" (AI100) entwickelt und wird jetzt als unabhängiges Projekt unter der Schirmherrschaft von AI100 am Human-Centered AI (HAI) Institut der Stanford Universität, CA, USA durchgeführt. Zu den Projekt Partner gehören Stanford, Harvard, MIT, SRI International, Partnership On AI, OpenAI und das McKinsey Global Institute. Bei Google Scholar h5-Index bedeutet ein fehlender Wert, dass wir keinen Eintrag gefunden haben.

*Wissensrepräsentation und Schlussfolgern*

Das Wissen über die Welt wird so dargestellt, dass eine Maschine es nutzen kann, um komplexe Aufgaben zu lösen, wie z.B. einen Dialog in einer natürlichen Sprache zu führen.

| **Wissensrepräsentation & Schlussfolgern**<br>Logik, Schlussfolgern, Wissensrepräsentation, Planung, ... |
|---|
| **CORE A*** |
| Intern. Conf. on Autonomous Agents and Multi-Agent Systems (AAMAS) **h5 40** |
| Intern. Conf. on Automated Planning and Scheduling (ICAPS) **h5 27** |
| **CORE A** |
| International Semantic Web Conference (ISWC) **h5 41** |
| Extended Semantic Web Conference (ESWC) **h5 30** |
| Journal of Automated Reasoning **h5 24** |
| Theory and Practice of Logic Programming **h5 20** |
| International Conference on Automated Deduction (CADE) **h5 20** |
| Intern. Conf. on Logic for Programming, AI and Reasoning (LPAR) **h5 17** |
| European Conference on Logics and Artificial Intelligence (JELIA) **h5 -** |
| Automated Reasoning with Analytic Tableaux and Related Methods (TABLEAUX) **h5 -** |
| Conference on Theoretical Aspects of Rationality and Knowledge (TARK) **h5 -** |
| International Conference on Logic Programming (ICLP) **h5 -** |
| **CORE B** |
| International Journal of Approximate Reasoning **h5 40** |
| International Joint Conference on Rules and Reasoning (RuleML+RR) **h5 12** |
| International Conference on Case-Based Reasoning (ICCBR) **h5 11** |
| International Journal on Semantic Web and Information Systems **h5 -** |
| **Not Listed in CORE** |
| Semantic Web Journal (SWJ) **h5 32** |

Unsicherheit und Schlussfolgern

Maschinen müssen mit Unsicherheit umgehen können. Ihr Wissen ist nicht vollständig. Die Welt kann nicht komplett beobachtet werden. Der Eintritt von Ereignissen ist nicht mit Sicherheit festzustellen oder sogar vorauszusagen

| **Unsicherheit & Schlussfolgern**<br>Probabilistische Modelle, Entscheidungstheorie, ... |
|---|
| **CORE A*** |
| Conference on Uncertainty in Artificial Intelligence (UAI) **h5 30** |
| ACM Conference on Economics and Computation (EC) **h5 -** |
| **CORE A** |
| Autonomous Agents and Multi-Agent Systems **h5 25** |
| ACM Transactions on Economics and Computation **h5 -** |
| **CORE Unranked** |
| AAAI Conf. on AI and Interactive Digital Entertainment (AIIDE) **h5 17** |
| International Conference on Web and Internet Economics (WINE) **h5 16** |
| International Conference on Algorithmic Decision Theory (ADT) **h5 13** |
| International Conference on Probabilistic Graphical Models (PGM) **h5 -** |

## Maschinelles Lernen

Eine Maschine lernt aus Beispielen und kann die erlernten Muster und das erlernte Wissen auf neue Beispiele verallgemeinern. Die Maschine lernt also aus Erfahrungen ihr zukünftiges Verhalten zu verbessern.

| **Maschinelles Lernen** <br> (Un)überwachtes Lernen, Reinforcement Learning, Deep Learning, Data Mining, … |
|---|
| **CORE A*** |
| Neural Information Processing Systems (NIPS/NeurIPS) **h5 134** |
| International Conference on Machine Learning (ICML) **h5 113** |
| IEEE Transactions on Neural Networks and Learning Systeme (zuvor IEEE Transactions on Neural Networks) **h5 87** |
| ACM SIGKDD International Conference on Knowledge discovery and data mining (KDD) **h5 77** |
| International World Wide Web Conference (WWW) **h5 76** |
| ACM International Conference on Web Search and Data Mining (WSDM) **h5 55** |
| ACM SIGIR Conf. on Research and Development in Information Retrieval (SIGIR) **h5 52** |
| Conference on Learning Theory (COLT) **h5 44** |
| Machine Learning Journal (MLJ) **h5 38** |
| IEEE International Conference on Data Mining (ICDM) **h5 37** |
| **CORE A** |
| Journal of Machine Learning Research (JMLR) **h5 79** |
| IEEE Transactions on Knowledge and Data Engineering **h5 77** |
| Neural Networks **h5 56** |
| ACM Intern. Conf. on Information and Knowledge Management (CIKM) **h5 49** |
| International Conference on Artificial Intelligence and Statistics (AISTATS) **h5 43** |
| Data Mining and Knowledge Discovery Journal (DMKD) **h5 35** |
| Neural Computation **h5 34** |
| SIAM International Conference on Data Mining (SDM) **h5 33** |
| European Conference on Machine Learning and Knowledge Discovery in Databases (ECML PKDD) **h5 30** |
| European Conference on Information Retrieval (ECIR) h5 **26** |
| Pacific-Asia Conference on Knowledge Discovery and Data Mining (PAKDD) **h5 23** |
| **CORE B** |
| ACM Conference on Recommender Systems (RecSys) **h5 40** |
| International Joint Conference on Neural Networks (IJCNN) **h5 32** |
| Neural Processing Letters **h5 23** |
| Information Retrieval **h5 20** |
| International Conference on Artificial Neural Networks (ICANN) **h5 14** |
| International Conference in Inductive Logic Programming (ILP) **h5 -** |
| **CORE Unranked** |
| Asian Conference on Machine Learning (ACML) **h5 13** |
| International Conference on Learning Representations (ICLR) **h5 -** |

## Wahrnehmung und Sehen

Die Fähigkeit zur Verarbeitung visueller Information ist eine Grundbedingung für künstliche Intelligenzen. Mithilfe von Regeln und Algorithmen werden Bilder und andere sensorischen Eingaben verarbeiten, interpretiert und generiert.

| Wahrnehmung und Sehen |
|---|
| **CORE A*** |
| IEEE Conference on Computer Vision and Pattern Recognition (CVPR) **h5 188** |
| International Conference on Computer Vision (ICCV) **h5 124** |
| IEEE Transactions on Pattern Analysis and Machine Intelligence (TPAMI) **h5 118** |
| **CORE A** |
| European Conference on Computer Vision (ECCV) **h5 104** |
| International Journal of Computer Vision **h5 65** |
| Computer Vision and Image Understanding **h5 44** |
| **CORE Unranked** |
| Asian Conference on Computer Vision (ACCV) **h5 30** |

## Verstehen und Generierung von natürlicher Sprache

Maschinen müssen natürliche Sprache verarbeiten, verstehen und generieren können, um eine weitreichende Kommunikation zwischen Mensch und Computer per Sprache zu schaffen.

| Verstehen und Generierung von natürlicher Sprache |
|---|
| **CORE A*** |
| Meeting of the Association for Computational Linguistics (ACL) **h5 87** |
| **CORE A** |
| Conference on Empirical Methods in Natural Language Processing (EMNLP) **h5 76** |
| North American Association for Computational Linguistics (NAACL) **h5 56** |
| International Conference on Computational Linguistics (COLING) **h5 31** |
| Conf. of the European Chapter of the ACL (EACL) **h5 28** |
| Conference on Natural Language Learning (CoNLL) **h5 26** |
| **CORE B** |
| International Joint Conference on Natural Language Processing (IJCNLP) **h5 19** |

## Mensch-Maschine Interaktion

(Computer)programme müssen mit ihrer Umwelt interagieren. Ziel ist es, eine möglichst weitreichende Kommunikation zwischen Mensch und Computer zu schaffen.

| Interaktion |
|---|
| **CORE A*** |
| CHI Conference on Human Factors in Computing Systems (CHI) **h5 86** |
| ACM Conference on Ubiquitous Computing (UbiComp) **h5 52** |
| ACM Transactions on Computer-Human Interactions **h5 36** |
| **CORE A** |
| International Conference on Intelligent User Interfaces (IUI) **h5 27** |
| **Not Listed in CORE** |
| IEEE Transactions on Human-Machine Systems **h5 32** |
| AAAI Conference on Human Computation and Crowdsourcing (HCOMP) **h5 -** |

## Robotik

Design und Entwicklung von Robotern, die mittels KI-Algorithmen autonome mit der physischen Welt interagieren.

| Robotik |
|---|
| **CORE A*** |
| International Journal of Robotics Research **h5 61** |
| IEEE Transactions on Robotics **h5 54** |
| Robotics: Science and Systems (RSS) **h5 49** |
| **CORE A** |
| IEEE/RSJ International Conference on Intelligent Robots and Systems (IROS) **h5 54** |
| Robotics and Autonomous Systems **h5 49** |
| Journal of Field Robotics **h5 38** |
| International Conference on Field and Service Robotics **h5 15** |
| International Conference on Control, Automation, Robotics, and Vision **h5 11** |
| International Symposium on Robotics Research (ISRR) **h5 -** |
| **CORE B** |
| International Conference on Robotics and Automation (ICRA) **h5 75** |
| Autonomous Robots **h5 33** |
| European Workshop on Learning Robots (EWRL) **h5 -** |
| **Not Listed in CORE** |
| The International Journal of Robotics Research **h5 61** |
| IEEE Robotics & Automation Magazine **h5 30** |
| IEEE-RAS International Conference on Humanoid Robots (Humanoids) **h5 26** |
| IEEE Robotics and Automation Letters **h5 22** |
| Conference on Robot Learning (CoRL) **h5 -** |

## KI und Ethik

Beleuchtet die moralisch-ethische Dimension der KI. Wie sollen Mensch und Maschine künftig miteinander umgehen?

| KI und Ethik |
|---|
| **CORE Unranked** |
| AAAI/ACM Conference on AI, Ethics, and Society (AIES) **h5 -** |